\documentclass[onecolumn,showpacs,aps]{revtex4}
\usepackage{epstopdf}
\usepackage{graphicx}
\begin{document}
\title{How occasional backstepping can speed up a processive motor protein}
\author{Martin Bier$^{1}$ and Francisco J.\ Cao$^{2}$}
\affiliation{$^{1}$Dept.\ of Physics, East Carolina University, Greenville, NC 27858, USA}
\affiliation{ $^{2}$Departamento de F\'{i}sica At\'omica, Molecular y Nuclear, Universidad Complutense de Madrid, Avenida Complutense s/n, 28040 Madrid, Spain}
\date{\today}
\begin{abstract}
Fueled by the hydrolysis of ATP, the motor protein kinesin literally walks on two legs along the biopolymer microtubule.  The number of accidental backsteps that kinesin takes appears to be much larger than what one would expect given the amount of free energy that ATP hydrolysis makes available.  This is puzzling as more than a billion years of natural selection should have optimized the motor protein for its speed and efficiency.  But more backstepping allows for the production of more entropy.  Such entropy production will make free energy available.  With this additional free energy, the catalytic cycle of the kinesin can be speeded up.  We show how measured backstep percentages represent an optimum at which maximal net forward speed is achieved.
\end{abstract}
\pacs{05.40.-a, 87.16.Nn}
\maketitle

Processive motor proteins are among the tiniest engines known to man.  These proteins utilize the energy of ATP hydrolysis to literally walk along a biopolymer \cite{Howard}.  In a living cell they help maintain organization by transporting cargo, like organelles or vesicles filled with chemicals.

Already one and a half decade ago the stepping of the processive motor protein kinesin was made visible on the nanometer scale with optical tweezers \cite{Howard}.  Early communications \cite{Block1,Block2} reported that 5\% to 10\% of all steps of kinesin were backward.  But smaller fractions were described later on as methods and materials improved and better resolutions were achieved; Ref.\ \cite{Yanagida} gave 1/220 and Ref.\ \cite{CarterCross} gave 1/802.   Theoreticians have always been interested in backstep fractions as they can help verify stochastic models.  However, throughout the literature backstepping has implicitly and explicitly been seen as an occasional malfunction of the stepping motor protein.

\begin{figure}[htbp]
\centering{\resizebox{5.5 cm}{!}{\includegraphics{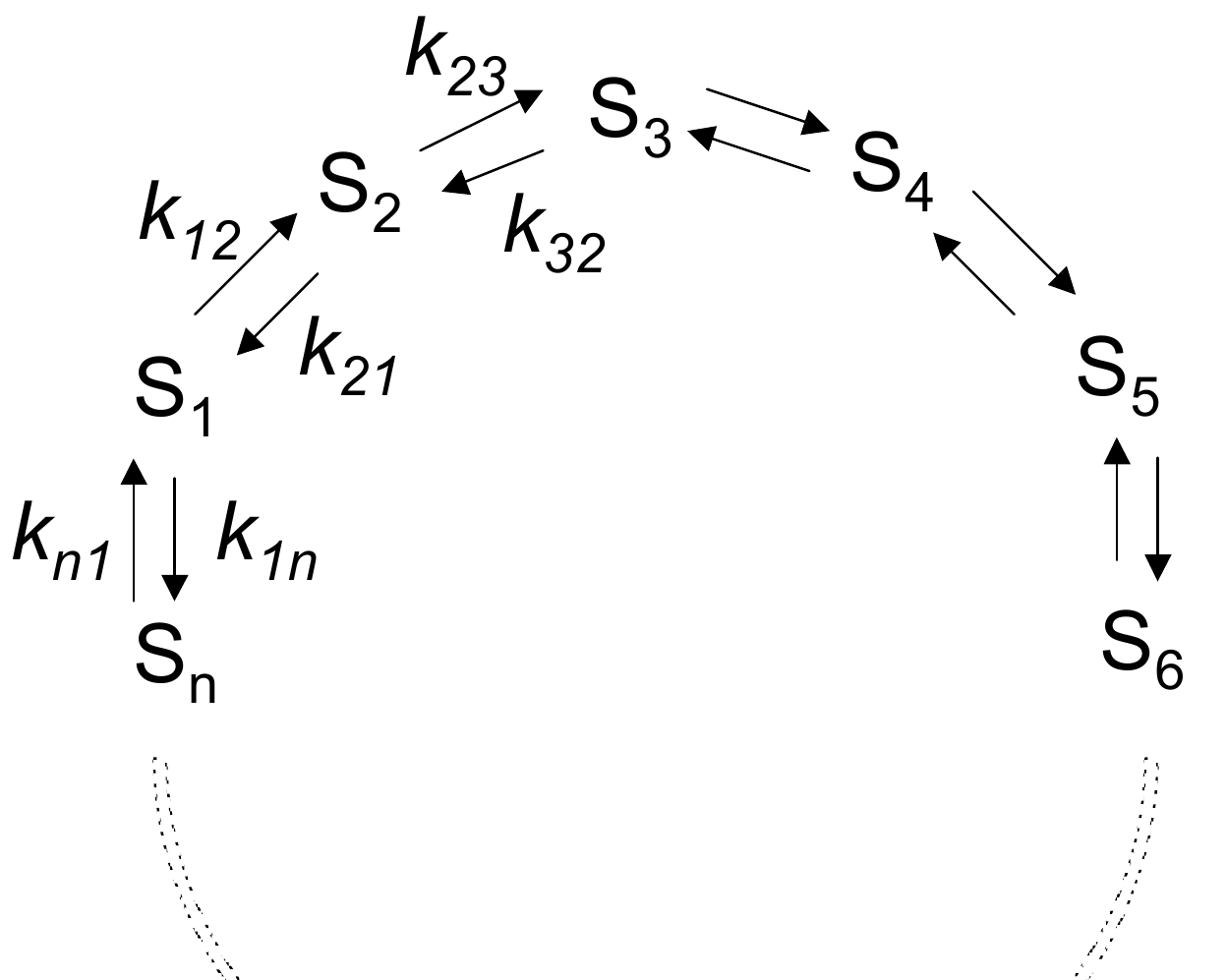}}}
\caption{An abstract conception of kinesin's catalytic cycle.  The cycle involves the binding and hydrolysis of ATP, the actual mechanical stepping, and the release of ADP and inorganic phosphate.  Each reaction is driven by an energy difference $G_{i,i+1}$ and the ratio of the forward and backward transition rate is an exponential function of $G_{i,i+1}$.}
\label{Fig1}
\end{figure}

In this Letter we will show how, in the Brownian environment of the motor protein, a ``well-tuned" backstep fraction can actually help the motor speed up.  We will show how the backstep fraction that leads to the highest net speed can be evaluated and how the resulting expression contains no freely adjustable parameters.  Finally, we will see how the experimentally established backstep fraction of kinesin is close to our predicted optimal backstep fraction.

The operation of an ion pump is generally modeled with a cycle as depicted in Fig.\ 1.  At equilibrium the product of the forward rates, $k_{12} \times k_{23} \times  ...\times k_{n1}$, equals the product of the backward rates, $k_{21} \times k_{32} \times  ... \times k_{1n}$, and no net cycling occurs.  To drive the protein through the sequence of states, $S_{1}, S_{2} \ ... \ S_{n}$, a driving energy is necessary \cite{Hill}.  Such energy comes available if one of the steps involves the binding of ATP and if the protein, in subsequent steps, catalyzes the hydrolysis of the bound ATP.  Eventually the remaining ADP and an inorganic phosphate have to be released so as to complete the cycle and to put the protein again in a state in which it can bind a new ATP.  Under physiological conditions the hydrolysis of ATP makes $G_{\rm ATP} = 22 \ k_{B} T$ units of free energy available.  In the course of a cycle of a membrane pump like Na,K-ATPase, part of $G_{\rm ATP}$ is utilized to bind, transport and release on the other side of the membrane three sodium ions and two potassium ions.  The transport is generally against the electrochemical potential of the involved ions.  Consistent with the model of Fig.\ 1, it is found that with a sufficiently low ATP-ADP potential and a high electrochemical potential for sodium and potassium the operation of the pump can be reversed \cite{lauger}.

\begin{figure}[htbp]
\centering{\resizebox{6 cm}{!}{\includegraphics{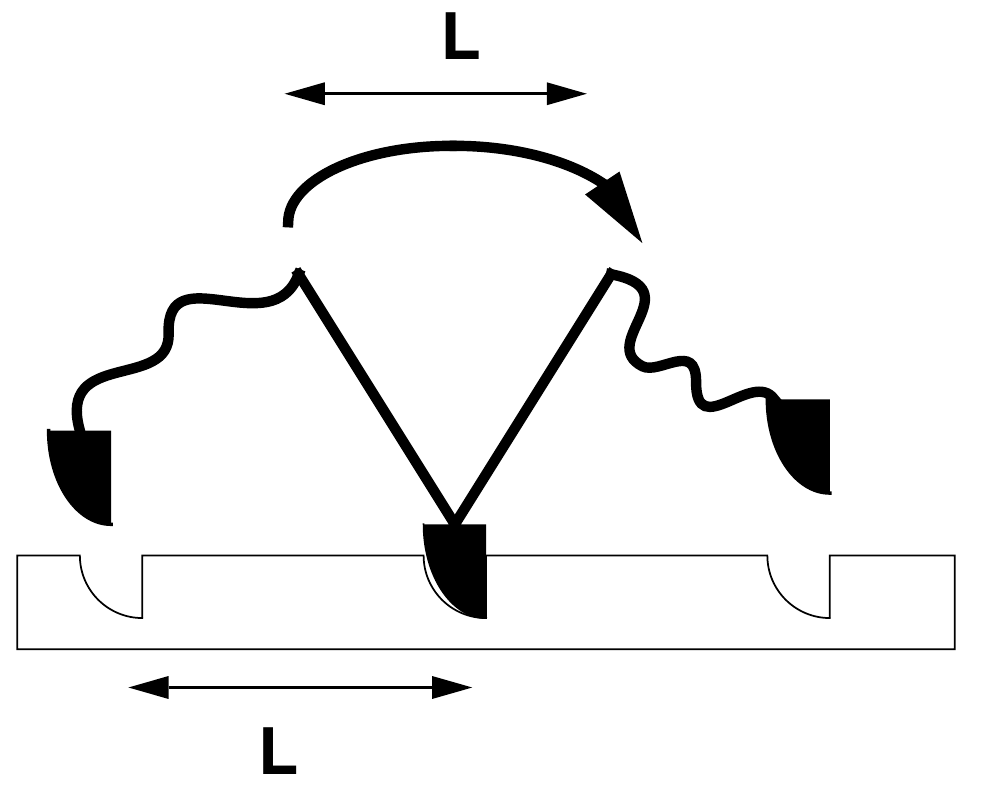}}}
\caption{The Brownian step of a processive motor protein.  After detachment of the trailing leg, the attached leg reorients and brings the detached leg to the vicinity of the next forward binding site.  After random diffusive motion will make the detached leg hit that binding site, actual binding can occur and a next step can commence.  A backstep occurs after rebinding at the rear site that the detached leg just came from.  The probabilities $p_{f}$ and $p_{b}$ for forward and backward binding depend on the energy behind the reorientation, the applied load force, and, as the process occurs in the Brownian regime, on $k_{B} T$, the natural unit of thermal energy.}
\label{Fig2}
\end{figure}

The stepping kinesin appears to operate in a fundamentally different way.  Tight coupling, i.e.\ an 8 nm step for every hydrolyzed ATP and a hydrolyzed ATP for every 8 nm step, has been observed for kinesin \cite{Block0,Gelles}.  Without a mechanical load it is just the $G_{\rm ATP}$ that is driving the cycle in Fig.\ 1.  Every mechanical step should correspond to one revolution around the catalytic cycle.  If a backstep would correspond to the cycle being run in the direction against the ATP hydrolysis, then we should have $p_{b}/p_{f} =  \exp [-G_{\rm ATP} ]$ for the ratio of the backstep probability and the  forward step probability.  However, $\exp [-22]$ turns out 7 orders of magnitude smaller than the measured backstep fractions that were mentioned before.   Furthermore, a model like in Fig.\ 1 leads to a stopping force, i.e.\ the load at which the kinesin comes to a standstill, that is determined by $G_{\rm ATP} = F_{st} L$, where $L$ is the steplength.  If $G_{\rm ATP} = F_{st} L$, then the two forces, chemical and mechanical, that are driving the cycle in opposite directions cancel each other out.  But, with $G_{\rm ATP} = 22$ and $L=8$ nm, the equation $G_{\rm ATP} = F_{st} L$ predicts a stopping force  $F_{st}$ that is about four times as large as the measured 7 pN \cite{Block1,Block2,CarterCross}.  Most importantly, it appears that kinesin still hydrolyzes ATP when it is pulled back with the stopping force and even when it is made to walk backwards with a load larger than the stopping force \cite{CarterCross,MolloySchmitz}.  All these observations make a model as depicted in Fig.\ 1 untenable.

We are thus led to a different model for the stepping motor protein \cite{BierPRL}.  After the detachment of the trailing leg, the attached leg reorients and brings the detached leg to the vicinity of the next forward binding site (see Fig.\ 2).  Brownian motion is eventually supposed to make the detached leg hit the next forward site.  Binding can then occur and a next step can commence.  However, the energies involved in this Brownian step are comparable to $k_{B} T$.  The energy $G$ that drives the reorientation of the attached leg is smaller than $G_{\rm ATP}$.  The detached leg can therefore rebind at the rear binding site that it came from and this can trigger the observed backstep.  We can thus arrive at $p_{b}/p_{f} \gg  \exp [-G_{\rm ATP} ]$ \cite{BierBioSystems}.  The scheme depicted in Fig.\ 3 is a more appropriate model for the Brownian stepper.  At a particular point in the ATP hydrolysis cycle a kind of coin-toss takes place and the forward-backward determination occurs.  The corresponding mechanical steps run in a dimension that is perpendicular to the plane of the chemical cycle.

\begin{figure}[htbp]
\centering{\resizebox{7.5 cm}{!}{\includegraphics{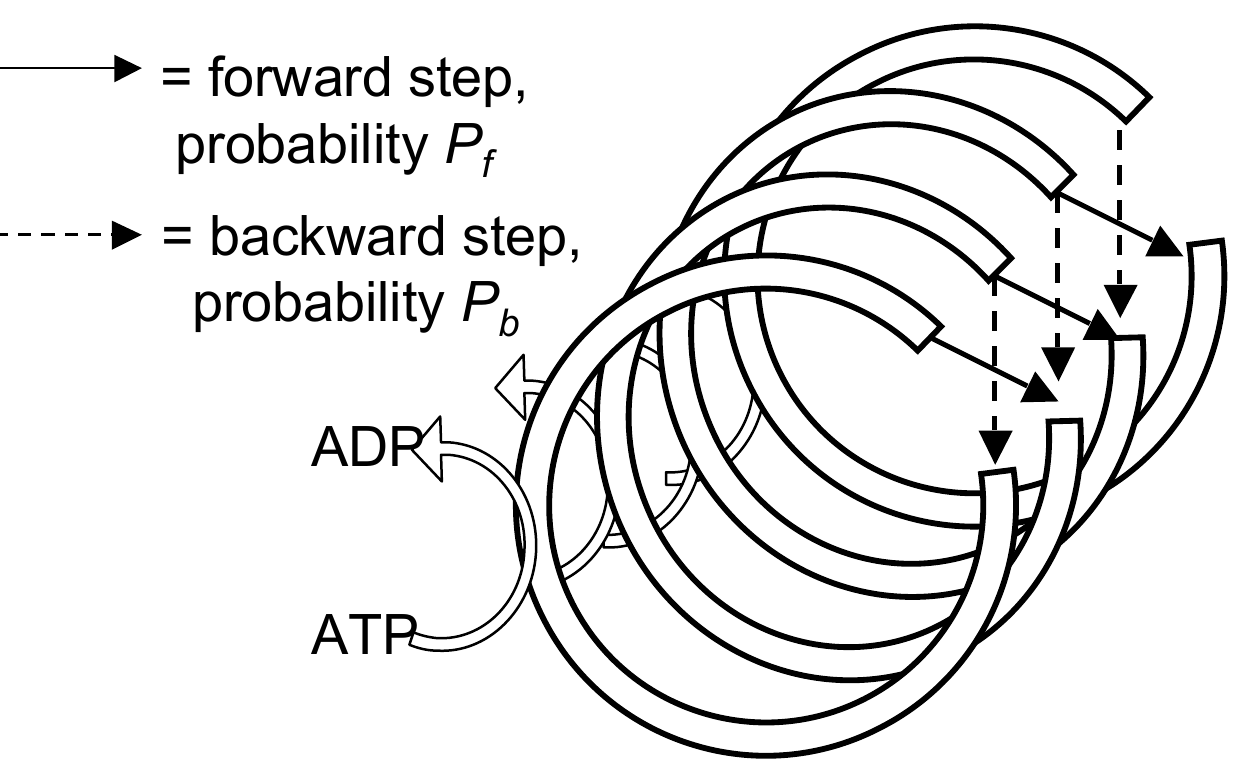}}}
\caption{Kinesin still hydrolyzes ATP when it is pulled back with a force larger than the stopping force.  A setup like Fig.\ 1, with the ATP-ADP potential and the mechanical load pushing a single cycle in opposite directions, is therefore not the appropriate model and needs modification.  Here we let the mechanical dimension run perpendicularly to the plane of the chemical cycle.  ATP hydrolysis drives the chemical cycle in a clockwise direction.  At the mechanical junction (cf.\ Fig.\ 2) a forward vs.\ backward ``decision" is made.}
\label{Fig3}
\end{figure}

The question that needs answering at this point is: why has natural selection led to a backstep probability that is many orders of magnitude larger than the bare minimum of $\exp [-22]$ that thermodynamics requires?  As was mentioned before, kinesin pulls organelles or chemical-filled vesicles across a eukaryotic cell.  The reason that eukaryotic cells have such an active transport system in the first place is that they are, unlike prokaryotic cells, too large to rely on diffusion for their transport needs.  Kinesin's speed ultimately determines how fast a eukaryotic cell can react to environmental stimuli.  There is an obvious selectional advantage in engaging a kinesin that runs faster.  No such straightforward evolutionary pressure for speedy turnover exists in the case of the aforementioned membrane pumps.  A cell or organelle can meet its need to extrude sodium ions by simply putting the right amount of Na,K-pumps in the membrane.

Our claim is that the entropy that is involved in the forward versus backward ``choice" can provide an answer to the question.  The free energy $k_{B} T \ln 2$ that is associated with a doubling of the available space is still small relative to the 22 $k_{B}T$-units released by ATP hydrolysis, but, we will argue, it can play a role in the optimization of a processive motor protein

Imagine a large number, $N$, of motor proteins that are all at the same position.  There is only one possible microstate for this macrostate.  Next, let all of these $N$ motors take a step.  If all of these motors step forward, we will continue to have only one microstate for the macrostate of the system.  But if, on the other hand, we allow for one backstep in the $N$ steps, then there are $N$ possible microstates.  This is because the backstep can occur with any of the $N$ motors.  The increase in the number of microstates implies an increase in entropy.  Another way to think of this is as follows.  Before a step occurs, let the entropy be $S_{0} = k_{B} \ln \Omega_{0}$, where $\Omega_{0}$ represents the number of microstates.  After one catalytic cycle and its accompanying step the number of microstates regarding the positions of the motors on the microtubules is $\Omega_{p}$.  The internal states of the motors are the same after the cycle as before, so the total number of microstates after the step is $\Omega_{1} = \Omega_{0} \Omega_{p}$.  For the ensuing entropy we have $S_{1} = k_{B} \ln \Omega = S_{0} + k_{B} \ln \Omega_{p}$.  The total increase in entropy is due solely to the increase in ``position entropy."  We thus have $\Delta S = - k_{B} \left( p_{f} \ln p_{f} + p_{b} \ln p_{b} \right)$ for the entropy increase that is associated with one step \cite{LeffRex}.

With a forward step probability $p_{f}$ and a backstep  probabilty $p_{b}$ we have for the speed of the stepper $v \propto p_{f} - p_{b} = 1 - 2 p_{b}$.  So the direct mechanical effect of an increase of the backstepping rate is a decrease of the speed, i.e.\ $\delta v \propto - 2 \ \delta p_{b}$.  But the backstep probability $p_{b}$ also makes free energy available to the amount of:
\begin{equation}
\tilde G = T \Delta S = - (p_{f} \ln p_{f} + p_{b} \ln p_{b} ) .
\label{tildeG1}
\end{equation}
Here $\tilde G$ is expressed in units of $k_{B} T$.  $k_{B} T$ can be thought of as the natural unit of thermal energy and we will express energy in $k_{B} T$-units throughout the remainder of this Letter.  As $p_{b}$ is observed to be small ($p_{b} < 0.01$), first order approximations will be sufficiently accurate and the theory we develop in this Letter is a first order theory in $p_{b}$.  At first order in $p_{b}$, Eq.\ (\ref{tildeG1}) reduces to $\tilde G \approx p_{b} (1 - \ln p_{b})$.  So a variation $\delta p_{b}$ in $p_{b}$ leads to a variation
\begin{equation}
\delta \tilde G \approx -( \ln p_{b}) \delta p_{b}
\label{deltatildeG}
\end{equation}
in the free energy that becomes available.

We contend that the $\tilde G$ of Eq.\ (\ref{tildeG1}) can be utilized to speed up the catalytic cycle of the motor protein.  Throughout biology, energy is used to build and maintain ordered, low entropy structures.  The sodium-glucose cotransporter concentrates glucose inside a cell.  It does so by using the energy of a cotransported sodium ion.  Entropic energy, in turn, is utilized whenever a transporter exploits a concentration gradient as a source of energy.  In this way there are many examples of biomolecules that use the generation of entropy as a source of free energy. In the course of its cycle the Na,K,2Cl-cotransporter brings one sodium ion, one potassium ion, and two chloride ions from the outside to the inside of a cell.  The cycle is electroneutral.  Under physiological conditions, this cotransporter effectively uses the transmembrane concentration gradients for sodium and chloride as an energy source to accumulate potassium inside the cell.

We next need to quantify how a $\delta \tilde G$ (cf.\ Eq.\ (\ref{deltatildeG})) can speed up a catalytic cycle.  To this end we return to the catalytic cycle depicted in Fig.\ 1.  The chemical part of kinesin's catalytic cycle (the binding of ATP, release of ADP and inorganic phosphate, detachment and attachment of the legs to the microtubule) is well modeled by a setup like Fig.\ 1.  With energy expressed in units of $k_{B} T$, the energy that drives the transition from $S_{i}$ to $S_{i+1}$ is $ G_{i,i+1}$ and we have $\exp [G_{i,i+1}] = {k_{i, i+1} / k_{i+1,i}}$ \cite{Howard}.  Adding $\delta \tilde G$ to $G_{i,i+1}$ we can achieve an increase $\delta k_{i,i+1}$ in the forward rate.
\begin{equation}
e^{G_{i,i+1}+\delta \tilde G} = {k_{i,i+1} + \delta k_{i,i+1} \over k_{i+1,i}} = {k_{i,i+1} \over k_{i+1,i}} \left(1 + {\delta k_{i,i+1} \over k_{i, i+1}} \right)
\end{equation}
For small $\delta \tilde G$ we have $\exp [ \delta \tilde G ] \approx 1 + \delta \tilde G$, which leads to
\begin{equation}
\delta \tilde G \approx {\delta k_{i,i+1} \over k_{i, i+1}}  \ .
\label{DeltaTildeG}
\end{equation}
So, ultimately, $\delta \tilde G$ is just the relative increase of $k_{i, i+1}$.  It is also possible to put the $\delta \tilde G$ towards reducing the reverse rate $k_{i+1,i}$.  However, when a reaction is already reasonably irreversible (i.e.\ $k_{i, i+1}/k_{i+1,i} $ being sufficiently large) this is not an efficient way to speed up the reaction.  We thus assume that the $\delta \tilde G$ brings down the energy of the product state and the energy of the activation barrier by the same amount.  In the following we will ignore $k_{i+1,i}$ and write $k_{i}$ for $k_{i,i+1}$.

The energies associated with the measured $p_{b}$'s are small.  For $p_{b} = 1/802$, Eq.\ (\ref{tildeG1}) gives $\tilde G = 0.01 \ k_{B} T$-units.  Following Eq.\ (\ref{DeltaTildeG}) we see that this translates into a 1\% change in the speed of a transition relative to $p_{b}=0$ case.  This may seem insignificant.  However, as we pointed out before, fast intracellular transport is an important, straightforward, and invariable selection criterion that is independent of the fitness landscape.

When one transition in a cycle like in Fig.\ 1 is speeded up by 1\%, it does {\em not} imply that the time to go through the entire cycle will decrease by 1\%.  That latter change will generally be less than 1\%.  We let $C^{v}_{k_{i}}$ be a so-called ``control coefficient" \cite{KacserBurns,HeinrichRapoport}; this is a dimensionless parameter that puts a figure on the influence that the rate $k_{i}$ of transition $i$ has over the cycling rate $v$ through the entire cycle.  This cycling rate is also the stepping speed of the motor protein.  We take
\begin{equation}
C^{v}_{k_{i}} = {\delta v / v \over \delta k_{i} / k_{i}}   \  .
\end{equation}
Going to the limit $\delta k_{i} \rightarrow 0$, we see that the control coefficient is essentially a logarithmic derivative, i.e. $\left( \partial \ln v \right) / \left( \partial \ln k_{i} \right)$.  So suppose $k_{i}$ is changed by 1\%.  The control coefficient $C^{v}_{k_{i}}$ then gives roughly the percentage by which $v$ changes as a result.  Assume that you change all of the transition rates in the cycle with the same percentage.  This is like scaling the time and, obviously, the speed $v$ would be changed by the same percentage.  This leads to the identity $\sum_{i=1}^{n} C^{v}_{k_{i}} =1$, which is commonly known as the Summation Theorem \cite{KacserBurns,HeinrichRapoport}.  If $k_{i}$ represents the rate limiting transition, then $C^{v}_{k_{i}}$ will be close to unity.  $C^{v}_{k_{i}}$ will be close to zero if the $k_{i}$-transition is very fast compared to other transitions in the cycle.  Generally, the control coefficient $C^{v}_{k_{i}}$ will be a positive number between zero and unity that expresses the amount of control that step $i$ has over the speed of the entire cycle.  In most biochemical networks and protein catalysis cycles there is no single rate limiting step.  Instead, control is somewhat equally distributed over several steps.

Since $\delta \tilde G$ already denotes the relative change of $k_{i}$, we have for the speed of the motor protein and its increase due to the addition of $\delta \tilde G$ to step $i$
\begin{equation}
v + \delta v \propto 1 + C_{k_{i}}^{v} \delta \tilde G \approx  1- C_{k_{i}}^{v} ( \ln p_{b}) \delta p_{b} \ .
\end{equation}
We combine this result with the mechanical effect of the backstepping, which, we saw earlier, was described by $v \propto 1-2p_{b}$ or, consequently, $\delta v \propto - 2 \ \delta p_{b}$.  We now have for the net variation of $v$ when $p_{b}$ is varied:
\begin{equation}
\delta v \propto - \left\{ 2 + C_{k_{i}}^{v}  \ln p_{b} \right\} \delta p_{b} \ .
\label{NetVariation}
\end{equation}
The speed $v$ has a maximum when the variation $\delta v$ equals zero, i.e.
\begin{equation}
\ln p_{b} = - 2/C_{k_{i}}^{v} \ .
\label{optimum}
\end{equation}
This equation relates the optimal backstepping fraction to the control coefficient of the transition that includes the free energy derived from the forward-versus-backward entropy.  The $p_{b}= 1/220$ from Ref.\ \cite{Yanagida} leads to $C_{k_{i}}^{v} = 0.4$.  The $p_{b}= 1/802$ from Ref.\ \cite{CarterCross} leads to $C_{k_{i}}^{v} = 0.3$.

Under physiological conditions the transitions in the cycle of a motor protein are driven by energies of about 2 $k_{B} T$-units \cite{CrossTrends,BierBioSystems}.  Consequently, reverse transitions in the catalytic cycle are relatively rare.  If we ignore the backward rates in the cycle of the motor protein, then the control coefficient $C_{k_{i}}^{v}$ is simply the fraction of time during the cycle that the motor protein spends in state $S_{i}$.  That time equals $1/k_{i,i+1}$.  When the back leg of kinesin detaches, the hydrolysis of the ATP bound to that leg proceeds concurrently.  The release of the ADP is supposed to take place upon rebinding of the detached leg to the microtubule.  In a comprehensive review \cite{CrossTrends} a rate of $\sim 250 \ {\rm s}^{-1}$ is given for the back-leg-detachment/hydrolysis transition.  With the stepping rate of $\sim 100 \ {\rm s}^{-1}$, we are indeed led to a control coefficient of 0.4 for this transition.  The same reference lists $\sim 300 \ {\rm s}^{-1}$ as the rate for the ADP release transition.  This would correspond to a control coefficient of about 0.3.  Both numbers are consistent with the backstepping rates of Refs.\ \cite{Yanagida} and \cite{CarterCross} and Eq.\ (\ref{optimum}) and its underlying hypothesis.

There is no specific biomolecular mechanism that we can point at for the utilization of the free energy that is made available by a slightly increased $p_{b}$ and, in that sense, a solid underpinning of our hypothesis is lacking.  It is generally true that having more configurations available within a macrostate effectively lowers the free energy of that state.  An example illustrates how this is often important on the level of a single biomolecule.  For a protein there are many ways to be unfolded and randomly coiled, i.e., there are many, say $\Omega_{u}$, unfolded microstates.   The number of microstates, $\Omega_{f}$, for the correctly folded state is generally orders of magnitude smaller \cite{Jackson}.  So the folding of a protein actually requires a free energy $T \Delta S = k_{B} T \ln (\Omega_{u} / \Omega_{f})$.  By making the folded state more flexible, i.e.\ by giving it more ``statistical wiggle room," this required free energy may be brought down and that may lead to faster folding.  To a large extent, the situation is analogous for the processive motor protein. By putting more available microstates at the end of the ATP hydrolysis cycle the motor can increase the free energy that is available to drive the step.  Molecular dynamics simulations have been widely used to shed light on the dynamics of protein folding.  They may become similarly illuminating for the case of stepping motor proteins.  What we have shown in this Letter is that for kinesin the measured percentage of backsteps and the measured rates for the conformational changes agree with the idea of occasional backstepping as a source of energy and a way to speed up the motor protein.

We are grateful to various people from the Humboldt Universit\"at in Berlin for critical feedback during two seminars there.  Figure 3 was an idea of Wolfram Liebermeister.  FJC acknowledges financial support from the MEC (Spain) through Research Project FIS2006-05895.

\end{document}